# Alzheimer's Disease Classification in Functional MRI With 4D Joint Temporal-Spatial Kernels in Novel 4D CNN Model

Javier Salazar Cavazos[1]   and Scott Peltier[2]
[1]Electrical & Computer Engineering, University of Michigan, Ann Arbor, MI, United States
[2]Biomedical Engineering, University of Michigan, Ann Arbor, MI, United States

## Synopsis

**Motivation:** Previous works in the literature apply 3D spatial-only models on 4D functional MRI data leading to possible sub-par feature extraction to be used for downstream tasks like classification.

**Goal(s):** In this work, we aim to develop a novel 4D convolution network to extract 4D joint temporal-spatial kernels that not only learn spatial information but in addition also capture temporal dynamics.

**Approach:** We apply our novel approach on the ADNI dataset with data augmentations such as circular time shifting to enforce time-invariant results.

**Results:** Experimental results show promising performance in capturing spatial-temporal data in functional MRI compared to 3D models.

**Impact:** The 4D CNN model improves Alzheimer's disease diagnosis for rs-fMRI data, enabling earlier detection and better interventions. Future research could explore task-based fMRI applications and regression tasks, enhancing understanding of cognitive performance and disease progression.

## Introduction

Resting-state functional MRI (rs-fMRI) is increasingly recognized as a biomarker for Alzheimer's disease (AD), with numerous studies reporting different blood-oxygen-level-dependent (BOLD) activations in specific brain regions relative to healthy subjects[1,2]. Feature extraction from this neuroimaging data typically involves machine learning algorithms applied to either functional connectivity matrices or subcortical surface maps[3,4]. Contrarily, this study focuses on the 4-dimensional data for classification which is often overlooked due to bigger computational demands. We evaluate three deep learning approaches for handling 4D data. The first approach employs a 3D convolutional neural network (CNN) using the ConvNeXt architecture[5], treating time samples as input channels. The second approach is a hybrid model combining a 3D CNN with a long short-term memory (LSTM) module[6] to separately capture spatial features and temporal dynamics. The third approach introduces a novel 4D CNN model that performs convolutions using 4D temporal-spatial kernels. While using a 4D CNN is not entirely unprecedented[7,8], this study represents the first application of such a model in rs-fMRI for diagnosing Alzheimer's disease.

## Methods

We used the Alzheimer's Disease Neuroimaging Initiative (ADNI) dataset[9], selecting 3T rs-fMRI scans (EPI pulse sequence data) with spatial resolution 3.3mm and a repetition time of 3000ms, comprising 140 temporal samples of size 65×77×65. Additionally, structural MRI scans were used to assist in the normalization process. Recognizing the usually limited size of medical datasets, we augmented the dataset by considering each session as an independent "pseudo-subject" to mitigate class imbalance and increase dataset size. To prevent cross-contamination between train and test sets, scans from the same individual were assigned exclusively to one set, resulting in class distributions of CN (602/50), MCI (210/50), and DAT (147/50)} for train/test samples. The test set was balanced to ensure accuracy is a meaningful metric and the validation set consisted of a subset of the train set through k-fold cross-validation.

Data preprocessing involved several steps: converting raw DICOM files to the BIDS format[10] and processing with fmriprep[11]. For structural scans, this included N4 bias field correction, skull stripping, and spatial normalization to the MNI152 linear space from TemplateFlow[12]. For functional scans, this entailed slice-timing correction, head-motion estimation, and fieldmap-less susceptibility distortion correction. Further preprocessing involved bandpass filtering between 0.01-0.1 Hz using scikit-learn[13], discarding the first 20 temporal samples, and applying Z-score normalization on each voxel's time series data.

All models were implemented using PyTorch, with data importation facilitated by the NiBabel[14] package. We addressed class imbalance using a weighted cross-entropy loss function with inverse frequency weights w=[959/602, 959/210, 959/147] and used the Adam optimizer[15] with weight decay and a cosine decay learning rate scheduler. Training was conducted on a system equipped with a 32-core CPU, 187GB RAM, and 1 NVIDIA 4090 24GB-VRAM GPU.

For the 4D CNN model, custom "Conv4D" layers were developed and integrated into our 4D convolutional blocks as illustrated in Figure 1. For the 3D CNN + LSTM model, spatial features for each time sample were separately extracted and globally averaged. Then, all of the collected time samples were provided to the LSTM module to be used for classification as shown in Figure 2. For the 3D CNN model, all time samples were treated as separate channel inputs.

## Results

The 4D CNN model better predicted patient diagnosis compared to other models as indicated in Table 1. For model interpretability, two analyses were conducted. First, the 4D kernels in the first layer were plotted against time to visualize features learned by the model (Figure 3). From this, it appears that some lower-level features include derivative and averaging information. Secondly, the Grad-CAM++ method[16] was employed to identify important regions used for diagnosis. Figure 4 presents the BOLD response in the hippocampus with corresponding Grad-CAM signal, and to the right, spatial saliency maps at a fixed time point, highlighting significant spatial features such as cerebellum, prefrontal cortex, and hippocampus.

## Discussion & Conclusion

This study demonstrates that diagnosis can be predicted using joint temporal-spatial kernels, illustrating the efficacy of a 4D CNN. The model outperformed other methods that rely on more conventional modeling assumptions such as separate spatial and temporal learning modules. Moreover, saliency maps indicate relevant brain regions used for diagnosis.

Future research directions can focus on extensions to task-based fMRI data, where "stressing" various networks beyond the default mode network could provide more insight into cognitive performance and Alzheimer's diagnosis. Additionally, the model could be restructured to work on regression-based tasks, such as score prediction, rather than classification.

## Acknowledgements

This work was fully supported by NIH grant R21 AG082204.

## Figures

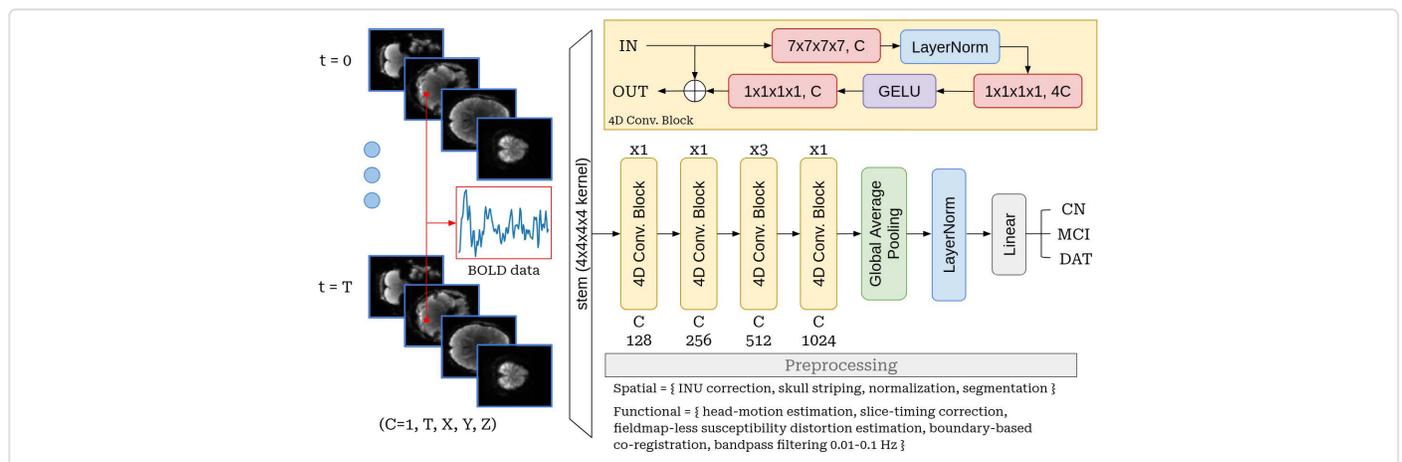

**Figure 1.** Proposed 4D CNN model architecture, consisting of four downsampling stages in a 1-1-3-1 configuration. The final stage outputs 1024 channels that are globally average pooled to yield 1024 features for the entire 4D scan.

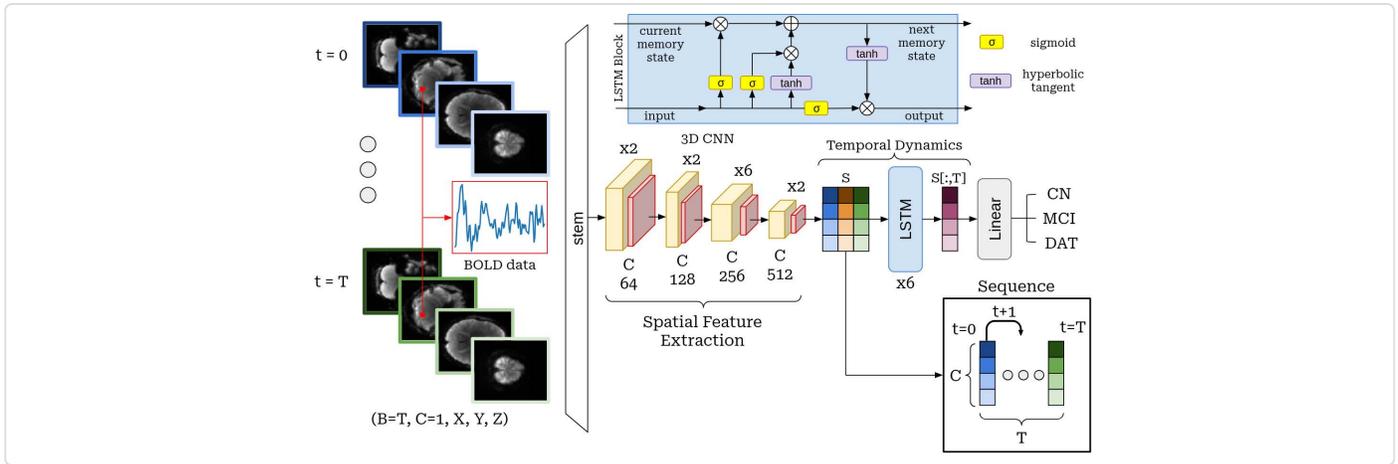

**Figure 2.** Architecture of the hybrid 3D CNN + LSTM model. Each 3D time sample is processed individually by the CNN, and the resulting features are aggregated into matrix S. The LSTM module captures temporal dynamics between time samples for classification purposes.

| Method | Accuracy | Sensitivity | Specificity |
|---|---|---|---|
| 2 class (CN/DAT) | | | |
| 3D CNN | 0.68 | 0.54 | 0.84 |
| 3D CNN + LSTM | 0.72 | 0.58 | 0.88 |
| 4D CNN | 0.77 | 0.62 | 0.92 |
| 2 class (CN/MCI) | | | |
| 3D CNN | 0.58 | 0.66 | 0.50 |
| 3D CNN + LSTM | 0.61 | 0.40 | 0.82 |
| 4D CNN | 0.63 | 0.64 | 0.62 |
| 3 class (CN/MCI/DAT) | | | |
| 3D CNN | 0.48 | 0.48 | 0.82 |
| 3D CNN + LSTM | 0.50 | 0.50 | 0.78 |
| 4D CNN | 0.53 | 0.53 | 0.83 |

**Table I.** Comparative results for the three approaches to handling the time dimension in raw 4D fMRI data. Accuracy, sensitivity, and specificity are reported for various class settings (binary and multi-class classification) using the ADNI test dataset.

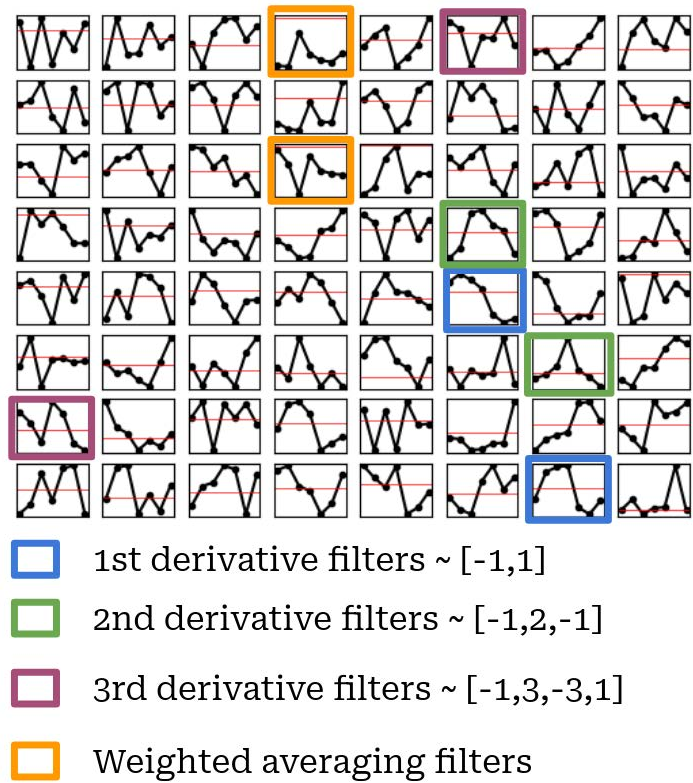

**Figure 3.** Temporal kernels from random spatial kernel locations for first layer channels (C=128). Only a subset of the total channels are shown for illustration simplicity. Moreover, only a few examples per filter are shown. The proposed model in the first layer extracts low-level features by using derivative and weighted average filters among other kernels less interpretable.

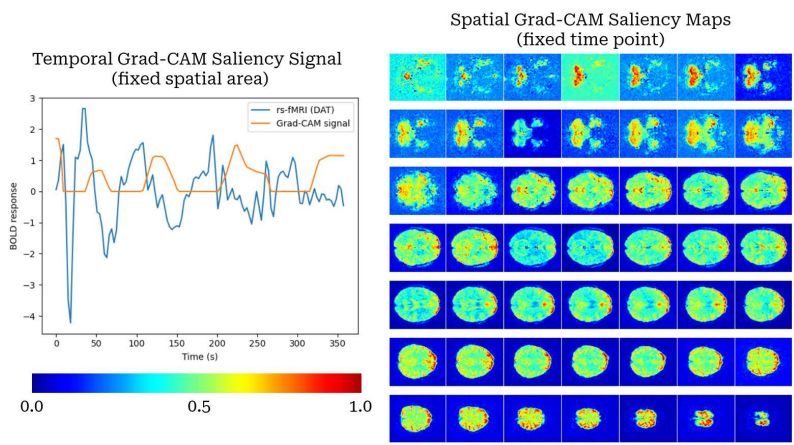

**Figure 4.** Model interpretability figure using the Grad-CAM++ method. Left image consists of the BOLD response at the hippocampus for a DAT diagnosed subject and corresponding Grad-CAM saliency signal over time. Right image consists of spatial Grad-CAM maps for a fixed time sample, illustrating key regions used for classification.